\documentclass[11pt,A4paper]{article}
\usepackage{graphicx}
\usepackage{amsmath}
\usepackage{amssymb}
\usepackage{epsfig}
\usepackage{natbib}    % reference
\usepackage[utf8]{inputenc}
\usepackage{authblk}

\usepackage{setspace}
\def\be{\begin{equation}}
\def\ee{\end{equation}}
\def\bea{\begin{eqnarray}}
\def\ena{\end{eqnarray}}

\def\blfootnote{\xdef\@thefnmark{}\@footnotetext} 
\long\def\symbolfootnote[#1]#2{\begingroup\def\thefootnote{\fnsymbol{footnote}}\footnote[#1]{#2}\endgroup} 

\title{Assessment and added value estimation of an ensemble approach  with a focus on global radiation forecasts}
\author[a,b]{Zied Ben Bouall\`egue}
\affil[a]{Deutscher Wetterdienst, Offenbach, Germany}
\affil[b]{Meteorological Institute, University of Bonn, Germany}
\date{}

\begin{document}

\maketitle

\begin{abstract}
The assessment of the high-resolution ensemble weather prediction system COSMO-DE-EPS is achieved with the perspective
of using it for renewable energy applications. 
The performance of the ensemble forecast %and of derived quantile products
is explored 
focusing on global radiation, the main weather variable affecting 
solar power production, and on quantile forecasts, key probabilistic products for the energy sector.
First, the ability of the ensemble system to capture and resolve the observation variability is assessed.
Secondly, the potential benefit of the ensemble forecasting strategy compared to
 a single forecast approach is quantitatively estimated.
%The decomposition of the quantile score on one hand and the ensemble interpretation of the continuous ranked probability score 
% on the other hand,  are used in order to propose a new metric called ensemble added value.
A new metric called ensemble added value is proposed, aiming at a fair comparison of an ensemble forecast with a single forecast,
when optimized to the users' needs.
Hourly mean forecasts are verified against pyranometer measurements 
over verification periods covering 2013. The results show in particular that 
the added value of the ensemble approach is season-dependent and increases with the forecast horizon. 

\end{abstract}

\section{Introduction}
The German electricity supply is currently being restructured,  aiming at increasing the integration of sustainable energies \citep{plan2020}.
Wind and solar energies are expected to play an important role in the ongoing energetic transition.
However, the intermittency of the power production, due to the weather dependent nature of these energy sources,
is a great challenge for the electricity grid operators \citep[e.g.,][]{boyle08}. 
High quality power forecasts  are thus required for management and operation strategies, 
to ensure efficiency and safety of the grid, as well as for energy trading.
In particular, the installed solar capacities in Germany are increasing rapidly and 
the attention paid to solar forecasting is growing simultaneously.
\\

The use of numerical weather prediction (NWP) models is a common approach for providing power forecasts 
for a horizon of a few hours to a few days \citep{costa2008,espinar2010}.
Weather forecasts are used as input for transformation models that deliver optimized power forecasts to the end users.     
The quality of the power forecasts therefore depends strongly on the quality of the underlying weather forecasts. 
NWP models find in this context new applications and efforts are being undertaken in order to improve the forecast quality
for weather variables relevant for the energy sector.
Forecasting  hourly photovoltaic power production
based on NWP models outputs has recently been  explored by \cite{lorenz2011} and \cite{zamo2014a}.
The deterministic power forecasts assessed in those studies show variability in skill 
over different seasons, weather situations and forecast horizons encouraging a probabilistic forecasting approach.
The limited predictability of weather events implies indeed a need for information about the predictive skill of the forecasts 
for an optimal use of the prediction systems \citep{krz83,richardson2000}
\\

Uncertainty about the  future state of the atmosphere can be estimated with an ensemble prediction system (EPS). 
An EPS provides a sample of possible future states of the atmosphere from multiple forecasts \citep{leut08}.
At Deutscher Wetterdienst (DWD) an operational cloud-resolving EPS that covers Germany has been running operationally since May 2012 \citep{gtpb10,pbtg12}.
The so-called COSMO-DE-EPS was originally designed to improve the quality of the forecast guidance in cases of high-impact weather events.  
Forecasts of convection-related weather events such as strong wind gusts and heavy precipitation can today be interpreted in a probabilistic way.
The range of applications of the system is now planned to be extended in order to support the integration of renewable energies into the German electricity grid. 
The performance of the ensemble system has therefore to be assessed focusing on energy relevant weather variables and relevant probabilistic products for the energy sector.
\\

The manuscript at hand deals with forecast verification of global radiation, the main weather variable
affecting solar power production. In terms of probabilistic products, the focus is on quantile forecasts that are key
products for energy applications \citep{pinson2007,morales2014}. Quantile forecasts are optimal point forecasts for users with an asymmetric loss
function \citep{gneiting2011}. In other words, users with different penalties associated with underprediction and overprediction 
can optimize their decisions by using quantile forecasts as decision variables.
\\

The quality of the ensemble forecasts and of the derived quantile products is estimated by means 
of proper scoring rules, namely the continuous ranked probability score \citep[CRPS,][]{crps2000} and the quantile score \citep[QS,][]{koenker99}.
The decomposition of proper scores provides an estimation of the  penalty due to the lack of reliability 
and  reward from resolution  \citep{brocker2009}.
Reliability measures the ability of the predictive distribution  to represent the unknown distribution of the observation conditional on the forecast while 
resolution measures the forecast's ability to distinguish between different subsets of observations  \citep{wilksv1,Bentzien2014}.
Since reliability can be corrected by  statistical techniques \citep{gneit2007}, %using training data from past forecasts and observations 
resolution, which is related to the forecast information content, is often considered as a more fundamental property which reflects the 'intrinsic value' of a forecasting
system \citep{toth2003}. 
\\

Besides the 'traditional' assessment of the ensemble forecast in terms of its statistical attributes (reliability
and resolution), the benefit of estimating the forecast uncertainty dynamically using an ensemble system is also quantified.
For this purpose, the ensemble forecast is compared to a single (deterministic) forecast.
Quantile forecast verification offers an appealing framework for a fair comparison of point forecasts in terms of potential performance.
Assuming that statistical adjustments can be applied similarly to any point forecasts, the comparison focuses on the forecast information contents.
The interpretation of the ensemble members in terms of quantiles allows the extension of the comparison to the whole probability
distribution described by the ensemble forecast. This approach enables the development of a new metric, which quantitatively estimates
the added value of the ensemble forecasting strategy compared to a single forecast approach.
\\

Results are shown for hourly global radiation forecasts from COSMO-DE-EPS against measurements from
32 pyranometer stations distributed over Germany. The results are discussed for verification periods of 90 days
in the year 2013. The ensemble added value is also shown and discussed as a function of the forecast horizon for
different periods of the year.
The manuscript is organized as follows: Section \ref{sec:data} describes the ensemble and observation datasets, Section
\ref{sec:verif} presents the verification methodology, Section \ref{sec:disc} discusses the results and Section \ref{sec:conc} concludes.

\section{Data}
\label{sec:data}
COSMO-DE EPS is a regional EPS run operationally at the German Weather Service (DWD). 
The ensemble system is based on a 2.8 km grid resolution version of the COSMO model \citep{stepp03,bald11}
with a model domain that covers Germany and part of the neighbouring countries. 
The ensemble comprises 20 members with variations in boundary conditions, model physics and initial conditions \citep{gtpb10,pbtg12}.
COSMO-DE-EPS has been developed focusing on high impact weather events.
Previous studies have discussed the performance of the system in this context \citep{bbtg2013,zbb2013}.
\\

Global radiation, the sum of direct and diffuse radiation, is here the model output parameter of interest.
Global radiation is defined as the total downward solar radiation (or irradiance) incident on a horizontal surface \citep{badescu}.
The performance of hourly mean ensemble forecasts  from the 0300
UTC run is explored for the year 2013. During this year, the forecast horizon of the operational ensemble system was 21
hours until March, and then extended to 27 hours.  Forecasts associated with solar zenith angles with cosine lower than 0.15 radians 
are considered as 'night' hours and are excluded from the verification process.   
Each forecast lead time is investigated separately. We mainly focus on a forecast horizon of 9 hours (valid at 0012 UTC) 
corresponding approximatively to the daily peak of solar power production. 
In this way, the strong diurnal cycle associated with solar variables does not affect the  interpretation of the verification results \citep{hamill2001}.
The impact on the intrepretation of the results due to the natural annual cycle is alleviated by using verification windows of 3 months.  
\\

\begin{figure*}
\centering
\includegraphics[width=8cm]{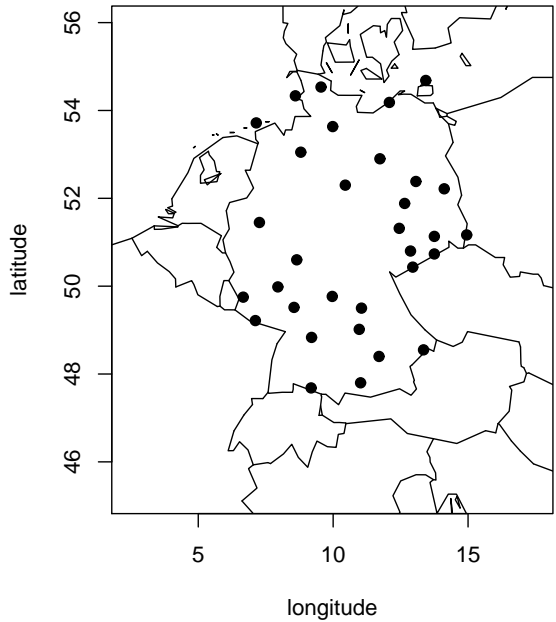}
\caption{
  Map of Germany {(approximately the model domain)} with latitude/longitude axes. Locations of the 32 pyranometer stations used in this study are shown. 
}
\label{fig:pyr}
\end{figure*}

\begin{figure*}
\centering
\includegraphics[width=8cm]{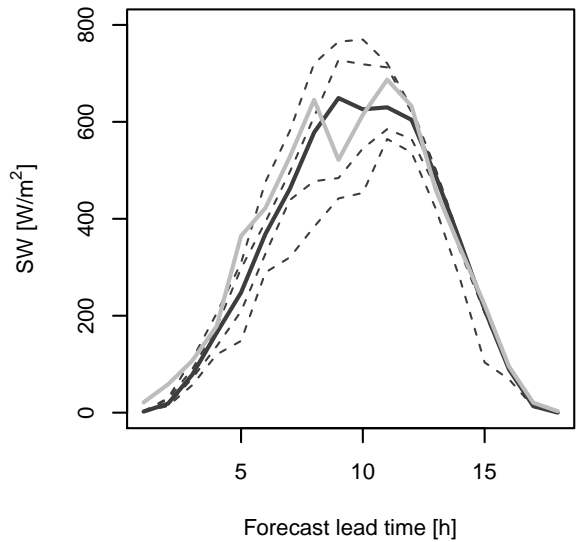}
\caption{
\protect{
  Global radiation COSMO-DE-EPS ensemble derived forecasts valid on July 5 2013 at Hamburg. The full grey lines correspond to the pyranometer measurements, the full black lines to the ensemble median (50\%-quantile) and the thin dashed lines indicate quantile forecasts for probability levels of 5\%, 25\%, 75\% and 95\% (from bottom to top).
}
}
\label{fig:exq}
\end{figure*}

Quality controlled pyranometer measurements are used for verification \citep{becker2012}. 
Figure \ref{fig:pyr} shows the geographical distribution over Germany of the 32 pyranometer stations used in this study.
For each observation point, the nearest model grid-point is selected. If not explicitly specified differently,  
quantile  forecasts are derived from the 20 ensemble members using a linear interpolation of the ensemble empirical cumulative density function.
An example of global radiation quantile forecasts from COSMO-DE-EPS valid at Hamburg on July 5 2013  is shown in Figure \ref{fig:exq}. 
Quantile forecasts with probability levels 5, 25, 50, 75 and 95\% are plotted as well as the corresponding observations.

\section{Verification methodology}
\label{sec:verif}
\subsection{General framework}

The aim is to predict global radiation at specific locations. % is aimed to be predicted.
The ability of COSMO-DE-EPS to achieve this task is questioned:    
are the forecasts able to capture the observation variability?
Are the forecasts able to do it in a valuable way, distinguishing between different subsets of observations? 
From this perspective, does the ensemble forecast provide additional information compared to a single member approach?
The assessment of the ensemble forecasts and of derived products is performed in order to answer these questions. 
Starting from well-established scores and their properties, suitable measures are proposed for this purpose. 
\\

The verification process is applied to the ensemble forecast as a whole as well as to derived quantile forecasts. A quantile forecast is defined by a nominal probability level: a quantile forecast of $q$ with nominal probability level $\alpha$ indicates that there is a probability  $\alpha$ that the observation will be less than $q$. So, in the following, a probability level can be interpreted in terms of the forecast cumulative probability distribution.
\\

The CRPS is a common tool for evaluating ensemble forecasts in the form of cumulative distributions. The sorted ensemble members are interpreted as quantile forecasts such as  
the ensemble describes a piecewise constant cumulative distribution function with jumps at the ensemble members \citep{crps2000}. Quantile forecasts at selected probability levels can be assessed separately. The QS is the natural tool used to evaluate such probabilistic products \citep{koenker99,gneiting2011b}. QS is based on the check function which can account for distinct costs for underprediction and overprediction.
\\

CRPS and QS are proper scoring rules \citep{gneitraf07,  Bentzien2014} and can be decomposed  \citep{crps2000,Bentzien2014}.
The decomposition produces reliability, resolution and uncertainty components of the scores. The decomposition of a proper score $S$ can be written as  
\begin{equation}
S= S_\text{reliability} - S_\text{resolution} + S_\text{uncertainty}.
\label{equ:dec1}
\end{equation} 
The reliability term reflects the forecast biases and the resolution term is related to the forecast information content.  The uncertainty terms depends only on the observations variability and is not influenced by the forecast \citep{wilksv1}.  
\\

The difference of the uncertainty and resolution parts is often denoted as the potential score \citep[e.g.][]{crps2000}: 
\begin{equation}
S_\text{potential}= S_\text{uncertainty}- S_\text{resolution} 
\label{equ:dec2}
\end{equation} 
$ S_\text{potential}$ measures the \textit{potential} performance in the sense that the reliability deficiencies can be potentially solved. Indeed, deficiencies in terms of reliability can be alleviated using past data and considering stationarity of the error characteristics. The statistical adjustment of forecasts based on past data is usually called \textit{calibration} \citep{gneit2007}. This step aims at providing reliable forecasts by correcting systematic biases and spread biases. Since the reliability term in Equation \ref{equ:dec1}  estimates the statistical consistency between the predictive distributions and the associated observations, $ S_\text{potential}$ can be interpreted as the score that should be obtained after statistical adjustment of the forecast. 
\\

In order to show the pertinence of using a given forecast, skill scores are computed. A skill score measures the relative benefit of using a forecast compared to a reference one \citep[e.g.][]{wilksv1}. We propose also to compute potential skill scores in order to compare forecasts and reference forecasts conditioned on calibration. In general terms, a skill score $Sk$ is defined as:
 \begin{equation}
Sk= 1-  \frac{S}{S^\star},
\label{equ:sk}
\end{equation}
where  $S^\star$ is the score of the reference forecast. % \citep{wilksv1}.
Applied to the CRPS and QS, Equation \ref{equ:sk} leads to the definition of the
continuous ranked probability skill score (CRPSS)  and of the quantile skill score (QSS), respectively.
Similarly, we define a potential skill score  $Sk_\text{potential}$  as:
\begin{equation}
Sk_\text{potential}= 1-  \frac{S_\text{potential}}{S^\star_\text{potential}}
\label{equ:skpot}
\end{equation}
where  $S^\star_\text{potential}$ is the potential score of the reference forecast. $Sk_\text{potential} $ measures the potential benefit of using a forecast compared to a reference forecast conditioned on calibration. Since the reliability terms of the forecast and of the reference forecast are not taken into account, the potential skill score
focuses on the forecast information contents.
\\

We consider in this study two different reference strategies: a climatology based forecast and a single forecast approach. Interpretations of \textit{skill score} and \textit{potential skill score} according to the chosen reference forecast is now discussed.

\subsection{Sample climatology as reference}

As a first reference forecast for the computation of skill scores, we consider the sample climatology.
For a given verification period, a cumulative probability distribution is derived from all the available observations.
This distribution and the related quantiles are then used as reference forecast over the verification period.   
\\

A climatological forecast, based on the observation sample, is by definition perfectly reliable \citep{crps2000}.
Moreover, a climatological forecast is a constant forecast and has therefore no resolution  \citep{mason04}. 
$S^\star$ and $S^\star_\text{potential}$ are equivalent and correspond to the uncertainty component of $S$:
\begin{equation}
S ^\star= S^\star_\text{potential}= S_\text{uncertainty}. 
\label{equ:climref}
\end{equation}
\\

The difference between potential skill score and skill score illustrates the benefit one can expect from calibration.
Considering the sample climatology as reference,
from Equations \ref{equ:dec2}, \ref{equ:sk}, \ref{equ:skpot}  and \ref{equ:climref} this difference is given as a simple ratio:
\begin{equation}
Sk_\text{potential} - Sk= \frac{S_\text{reliability}}{S_\text{uncertainty}}.
\label{equ:skdif}
\end{equation}
The difference between potential skill score and skill score emphasizes  the reliability deficiency relative to the observation variability.
A difference close to zero indicates a forecast able to capture  perfectly the observation variability.  
\\

From Equations \ref{equ:dec1}, \ref{equ:skpot} and \ref{equ:climref}, we can deduce that the potential skill score with climatology as reference corresponds to:
 \begin{equation}
Sk_\text{potential}= \dfrac{S_\text{resolution}}{S_\text{uncertainty}}.
\label{equ:skpotclim}
\end{equation}
$Sk_\text{potential}$ can  be interpreted as the proportion of observation variability that the forecast is able to correctly resolve. 
A potential skill score close to 1 indicates  perfect resolution of the forecast and
 a potential skill close to zero indicates no resolution at all. 
\\

Using the sample climatology as reference, reliability deficiency and resolution performance 
of the forecast are analyzed with respect to the variability of the observations.
Two important and complementary statistical aspects of the forecast quality, reliability and resolution,
can be discussed over different verification periods. The next step consists in evaluating how much of additional information is provided 
by the ensemble system compared to a single forecast. 
This step is taken considering a control forecast  (or by default an arbitrarily selected member of the ensemble)
as a reference forecast.

\subsection{Single forecast as reference}
\label{sec:verif3}

The CRPS reduces to the mean absolute error (MAE) when applied to a single (deterministic) forecast
 which allows a direct comparison of ensemble and deterministic forecast performance \citep{gneitraf07}.
However, from the user perspective, this comparison is often not relevant since two different types of forecasts are compared: a probability distribution and a point forecast. 
A forecast in the form of a probability distribution is usually  transformed into a probabilistic product adapted to the user's needs: 
a probability forecast associated with an event or a quantile forecast for a selected probability level. 
A comparison based on probability products is not suitable since the interpretation of a deterministic forecast in terms of probability
reduces its information content by transforming it into a binary outcome.
\\

A direct comparison of deterministic and ensemble derived forecasts  based on quantiles seems, on the other hand, adequate for including all the information content present.
Both types of forecasts, deterministic and quantile forecasts, are point forecasts, expressed as continuous variables when dealing for example with global radiation or wind speed. 
The step of interpreting a point forecast as a quantile does not deteriorate the forecast information content 
since it only requires the definition of a nominal probability level. 
This probability levels accounts for the user's sensitivity to underprediction and overprediction.
Any point forecast can potentially be adjusted to the user needs by calibration.    
A fair comparison of point forecast can then be performed focusing on the information content of the forecasts. 
\\

The quantile score (QS) and its decomposition offer an adequate framework for this comparison. 
QS can be applied similarly to any point forecast,
whether derived from an ensemble or a deterministic forecast, and
the QS decomposition provides an estimate of the forecast resolution. 
A fair comparison of ensemble derived quantiles and deterministic forecasts is then based on 
the potential quantile score which reflects the balance between 
observation variability (uncertainty) and forecast information content (resolution).
The reliability terms are disregarded on the assumption that calibration 
can be applied similarly to an ensemble forecast or to a deterministic forecast. 
Adequate statistical methods for such calibration exists, for example quantile regression \citep{koenker99}. However,  
the step of calibrating the forecasts, which requires sufficient past training data, is avoided here by 
relying on the decomposition of the quantile score. 
\\

QS is calculated for the derived ensemble forecast and for the reference forecast assigning the deterministic forecast to the relevant quantile.
The decomposition of the quantile scores allows then to extract a measure of the usefulness of the forecasts for the corresponding user. 
Components of the quantile scores are computed following \cite{Bentzien2014}.
We define  $PQS_{\tau}$  the potential quantile score of the $\tau$-quantile forecast derived from the ensemble forecast,    
and $PQS^\star_{\tau}$  the potential quantile score of the reference forecast for a probability level $\tau$.
%In this case, the probability level $\tau$  is used in the check function to penalized appropriately underprediction and overprediction of the ‘raw’ deterministic forecast. 
Following Equation \ref{equ:skpot}, the potential quantile skill score is computed as:
\begin{equation}
QSS_\text{potential} = 1- \dfrac{PQS_\tau}{PQS^\star_\tau} 
\label{equ:qsspot}
\end{equation}
Choosing a single forecast as reference, 
the potential QSS becomes a measure of the  added value of the ensemble system for a given probability level. 
\\

The added value of the ensemble forecasts considered as a whole is derived from the relationship 
between CRPS and QS.  It has been shown that 
the CRPS is equivalent to a weighted sum of the quantile scores applied to the sorted ensemble members \citep{broecker012}.
As already noted, the ensemble members are interpreted as quantile forecasts for the computation of the CRPS.
More precisely, the probability level $\tau_m$  corresponding to the sorted member of rank $m$ is defined as:
 \begin{equation}
\tau_m=\frac{m-0.5}{M}, \;\; m=1,...,M
\label{equ:tau}
\end{equation}
with $M$ the ensemble size. Noting $QS_{\tau_m}$ the quantile score at probability level $\tau_m$, 
the relationship between QS and CRPS is written as:
\begin{equation}
CRPS = \dfrac{2}{M} \sum_{m=1}^M QS_{\tau_m}.
\label{equ:crpsqs}
\end{equation}
Similarly, noting  $PQS_{\tau_m}$ the potential quantile score at probability level $\tau_m$, we define a potential CRPS as:
\begin{equation}
CRPS_\text{potential} = \dfrac{2}{M} \sum_{m=1}^M PQS_{\tau_m}
\label{equ:crpsqspot}
\end{equation}
$CRPS_\text{potential}$ reflects the potential performance of an ensemble forecast conditioned on the reliability of each sorted member. 
This assumption is stronger than in the computation of the potential CRPS following \cite{crps2000} where 
the reliability of the ensemble as a whole is considered. 
In this latter case, the reliability term of the CRPS decomposition is directly related to the rank histogram \citep{crps2000} 
and can therefore be subject to misinterpretation \citep{hamill2001}. 
\\

In the same way, potential quantile scores of the reference forecast are estimated for each probability level defined by the ensemble.
The associated potential CRPS is then defined as:
\begin{equation}
CRPS^\star_\text{potential} = \dfrac{2}{M} \sum_{m=1}^M PQS^\star_{\tau_m}
\label{equ:crpsqspot}
\end{equation}
where  $PQS^\star_{\tau_m}$ are the potential quantile scores at probability level $\tau_m$ when applied to the reference forecast. 
This procedure consists consequently in bringing the reference deterministic forecast to the degree of complexity of the ensemble forecast and not the opposite,
as it is the case, for example, when the ensemble mean is computed in order to be compared with a deterministic forecast.
\\

Finally, we define a new metric called ensemble added value (EAV) which takes the form of a potential skill score:  
\begin{equation}
 EAV = 1-  \dfrac{CRPS_\text{potential}}{CRPS^\star_\text{potential}}
\label{equ:skdif}
\end{equation}
EAV is a summary measure of the potential  benefit, conditioned on calibration, of using the ensemble forecast rather than a single forecast. 
%The range of  quantile forecast users uniformly distributed over the ensemble quantile levels.
An EAV greater than 0  indicates that the ensemble forecast outperforms the single forecast in terms 
of valuable information content.

\section{Results}
\label{sec:disc}

\subsection{Verification process}
\label{sec:res1}

COSMO-DE-EPS global radiation forecasts were assessed over the year 2013. 
Rolling verification windows were used in order to evaluate the performance for different periods of the year. 
The size of the verification windows is of 90 days and the rolling step of 10 days.
The results are first shown for forecasts from the 0300 UTC run  valid  at 0012UTC, corresponding to a forecast horizon of 9 hours.  
Verification results as a function of the forecast lead time are then  discussed for different seasons.
\\

The statistical significance of the results was  estimated by applying  bootstrapping.
Bootstrapping is  a common resampling technique proposed by  \cite{efrontib86}
and popularized in meteorology by \cite{hamill99}.
Confidence intervals of 5\% and 95\% are attributed to the scores,  derived from a 500-member block bootstrap sample. 
Each day is considered as a separate block of fully independent data in a way that the score distribution, from which the confidence intervals are drawn, 
represents the variability of the scores over the verification period and not between locations. 

\subsection{Reliability and resolution}
\label{sec:res2}
    
\begin{figure*}
\centering
\includegraphics[width=8cm]{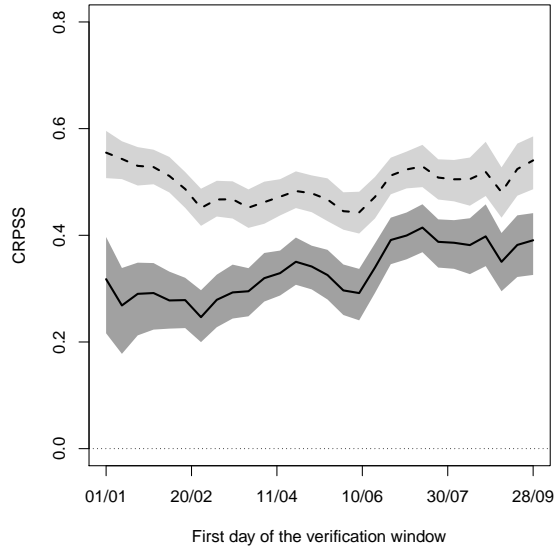}
\caption{
\protect{
CRPSS (full line) and potential CRPSS (dashed line) as a function of verification periods of 90 days in the year 2013. The skill scores are computed with the climatology as reference. The grey areas indicate the 5\% and 95\%  confidence intervals derived with bootstrapping. 
}
}
\label{fig:crpss}
\end{figure*}

\begin{figure*}
\centering
\includegraphics[width=4.5cm]{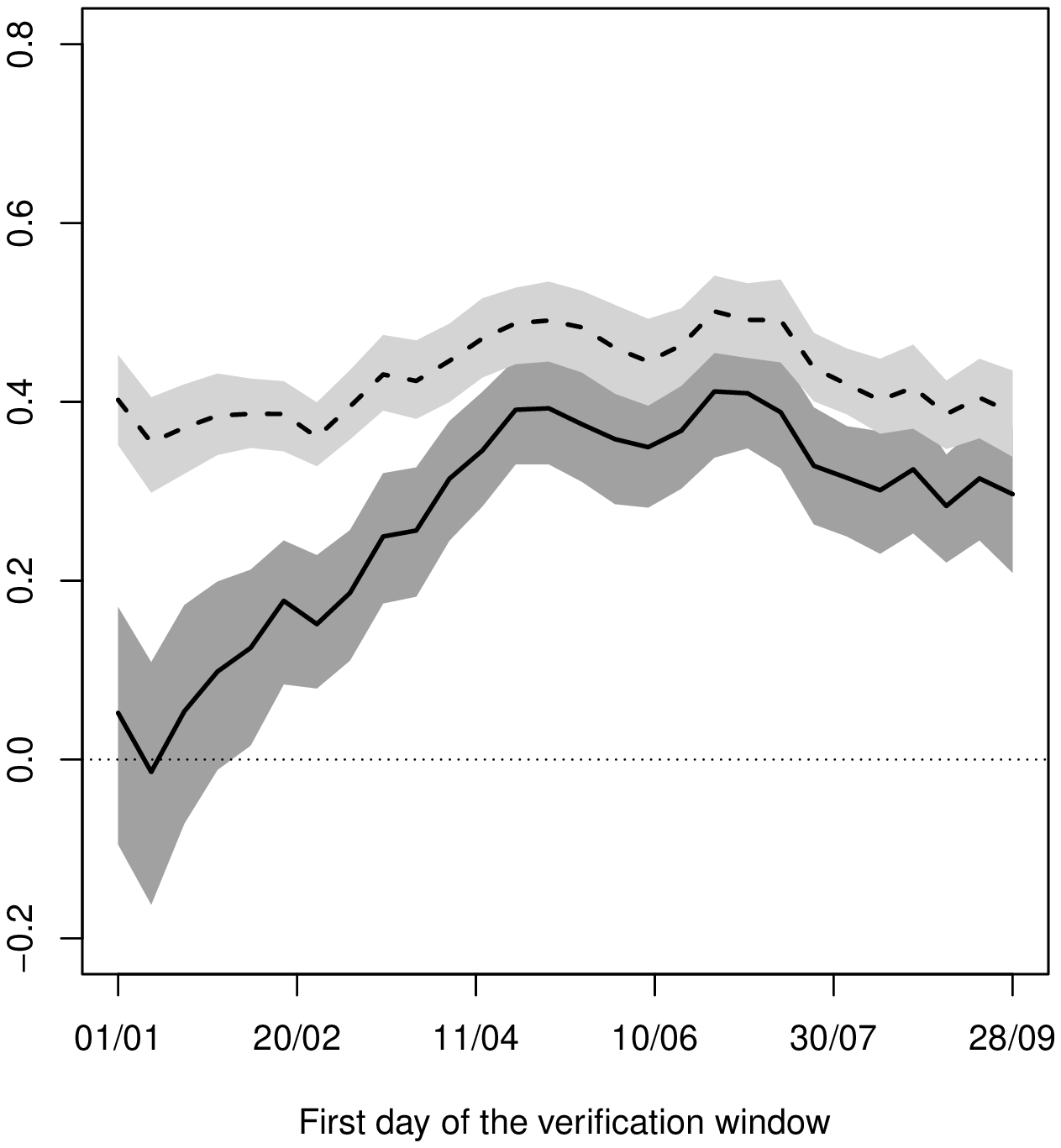}
\includegraphics[width=4.5cm]{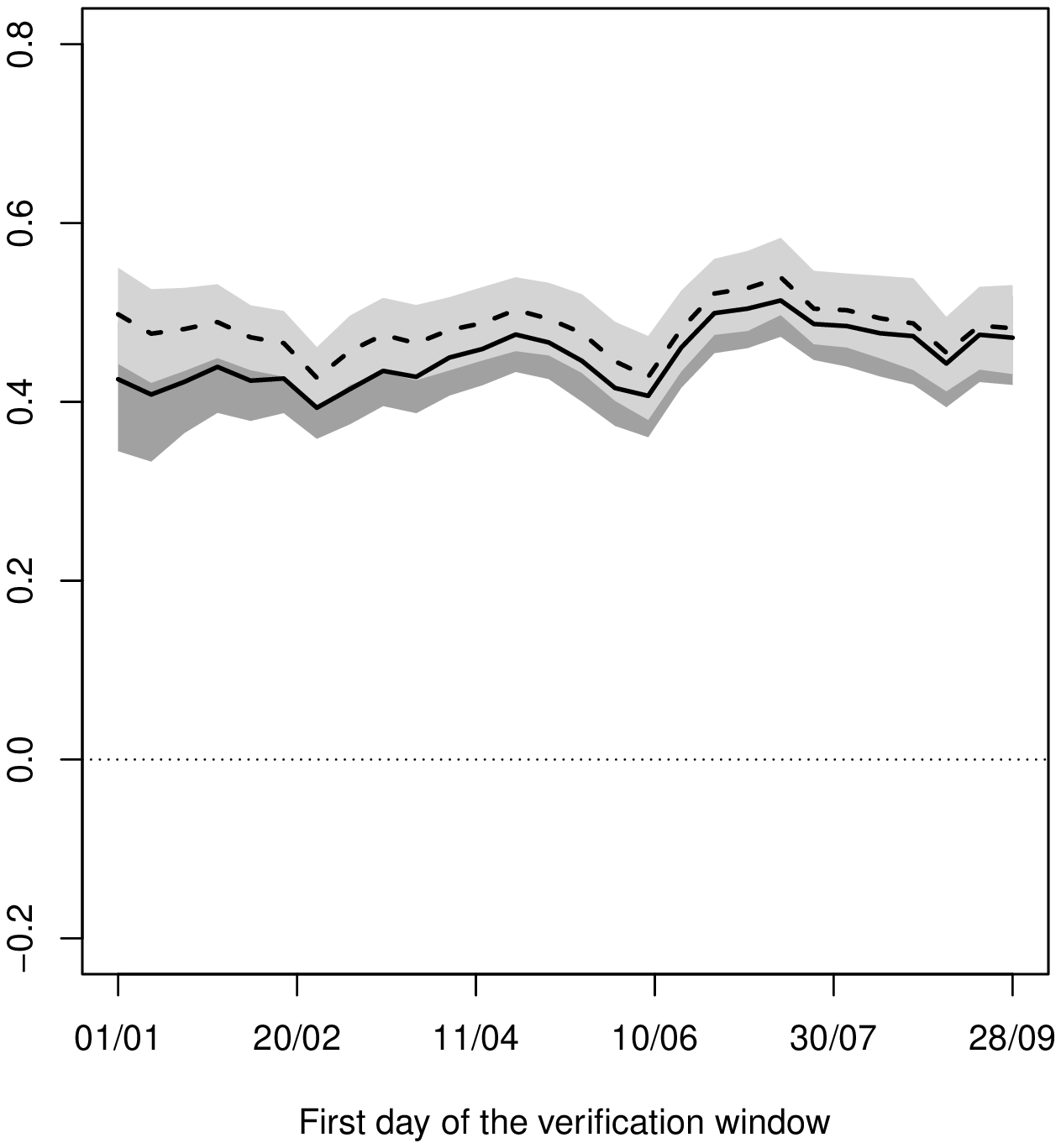}
\includegraphics[width=4.5cm]{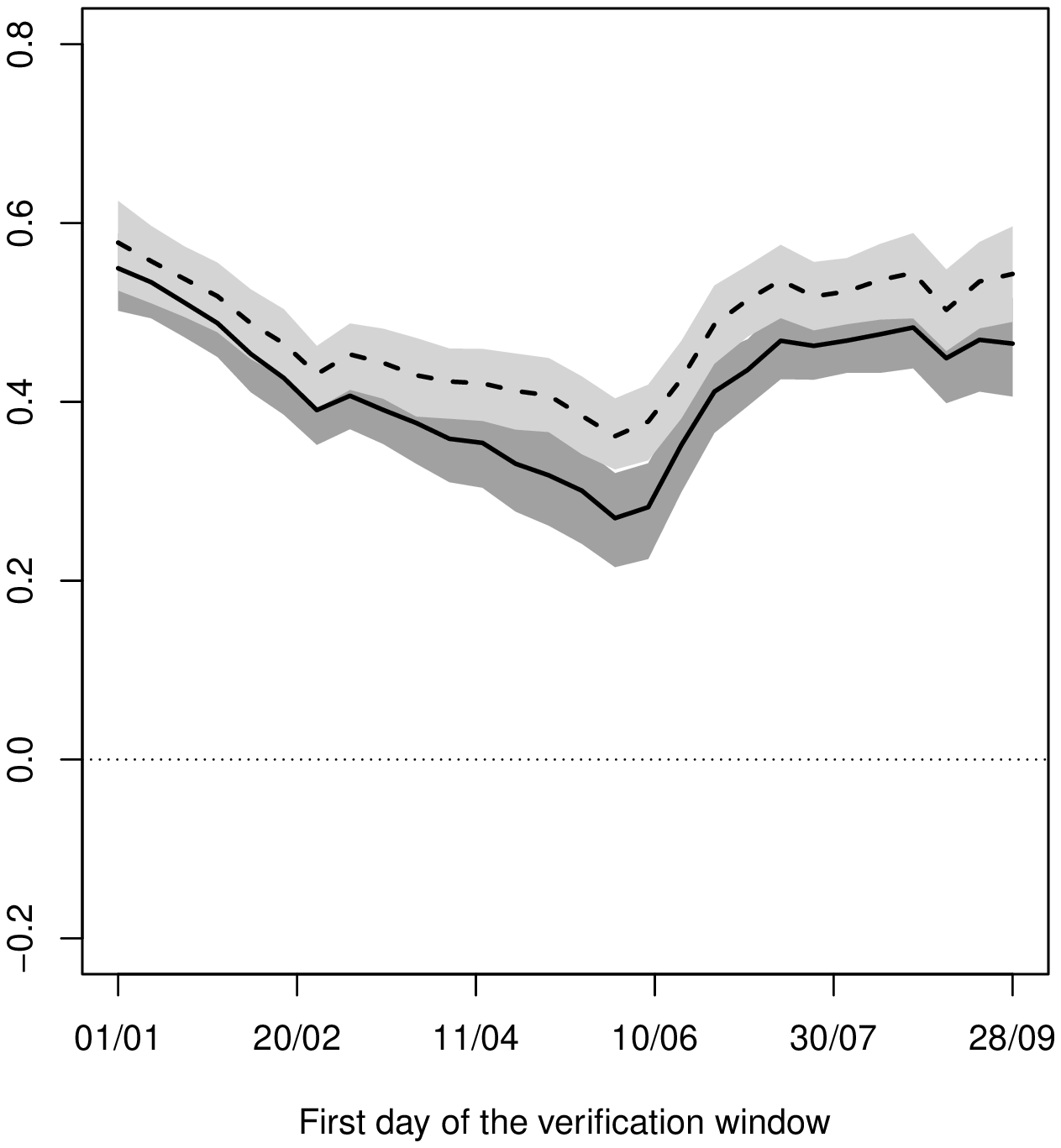}
\caption{
\protect{
QSS (full lines) and potential QSS (dashed lines) for the 25\%- (left), the 50\%- (middle) and the 75\%-quantile forecasts (rigth) as a function of verification periods of 90 days in 2013. The skill scores are computed using quantile climatological forecasts as reference. The grey areas represent the 5\%-95\% confidence intervals derived with bootstrapping. 
}
}
\label{fig:qss}
\end{figure*}

Skill scores of the ensemble forecast and of individual quantile forecasts are shown in Figures \ref{fig:crpss} and  \ref{fig:qss}.
The CRPSS and potential CRPSS with climatology as reference are plotted in Figure \ref{fig:crpss}. 
Similarly, QSS and potential QSS with climatological quantile forecasts as reference are shown in Figure \ref{fig:qss}.
The probability levels investigated corresponds to the median (50\%) and to the lower and upper quartiles (25\% and 75\%).
\\

The reliability of the ensemble forecasts is deduced from the difference between potential skill scores and skill scores:
a smaller difference means a more reliable forecast.
Reliability deficiencies occur throughout the year but are stronger during winter.  In particular,
the need for calibration is more pronounced when looking at the 25\%-quantile forecasts. 
The deficit of reliability at this probability level has a strong negative impact on the forecast skill.  
On the other hand, the median and upper quartile forecasts show correct statistical properties.  
Their good performance in terms of reliability is confirmed by the analysis of quantile reliability diagrams for different verification periods (not shown).   
\\

The ability of the forecasts to resolve the observed variability is described by the potential scores.
The potential CRPSS is not subject to large variations along the year, varying around  50\%. 
Similar results are obtained for the 50\%-quantile forecast.
The potential QS  for the two other probability levels present opposite tendencies as a function of the verification period: 
in terms of resolution ability, 
the 25\%-quantile forecasts have relatively higher skill in summer than  in winter while
the 75\%-quantile forecasts perform relatively better in winter than  in summer, compared to their relative climatological forecast.
This behavior can be explained by the fact that global radiation is a bounded variable. 
In summer, when clear days dominate,
the climatological distribution provides  a better (worse) estimation of  the upper (lower) quantiles 
than in winter, where cloudy days dominates, and vice versa. 
\\

In summary, the ability of the ensemble system to capture the observed variability is season dependent, while 
its ability to distinguish between subsets of events is relatively constant. At the quantile forecasts level,
reliability deficiencies have a severe impact on the forecast skill for small probability levels while  
the resolution ability at low and high probability levels exhibits a balancing effect between summer and winter periods.

\subsection{Ensemble added value}
\label{sec:res2}

\begin{figure*}
\centering
\includegraphics[width=8cm]{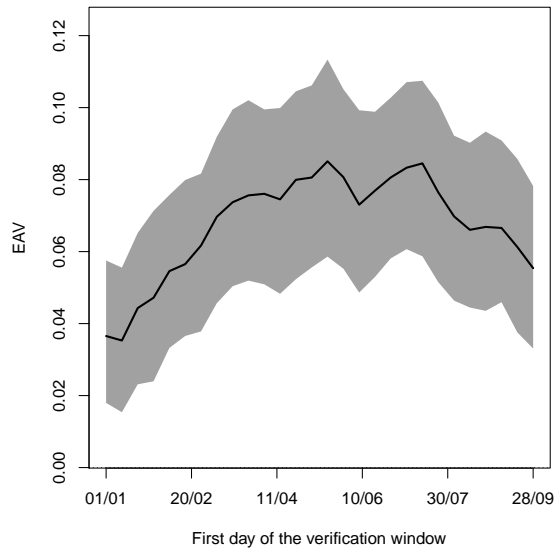}
\caption{ 
\protect{
EAV,  added value of the ensemble forecast strategy with respect to a single forecast approach, as a function of verification periods of 90 days in 2013. 5\%-95\% confidence intervals derived with  bootstrapping are represented by the grey area.
}
}
\label{fig:eav}
\end{figure*}

\begin{figure*}
\centering
\includegraphics[width=8cm]{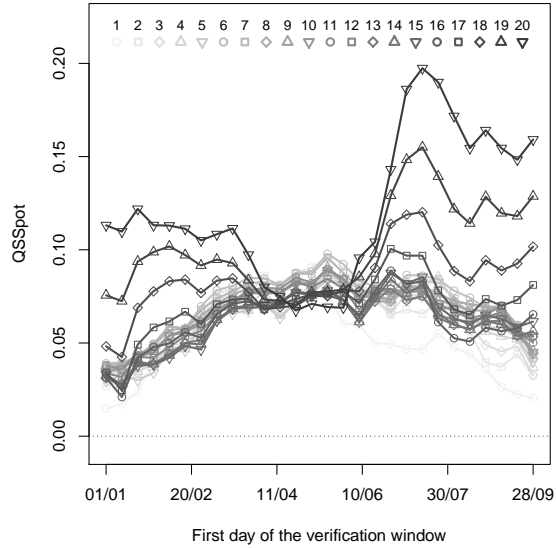}
\caption{ 
\protect{
  Potential QSS, with a single forecast as reference, as a function of verification periods of 90 days in 2013. Results for probability levels 2.5\%, 7.5\%,..., 92.5\%, 97.5\% (cooresponding to the ensemble members of rank 1, 2, ..., 19, 20) are represented with different shades of grey.
}
}
\label{fig:qav}
\end{figure*}

The benefit of the ensemble strategy is now illustrated and discussed.
For each verification day, one member from the 20 ensemble members was randomly selected  and  
used as a reference forecast.  The EAV estimated this way is shown in Figure \ref{fig:eav}.
Note that similar results are obtained if
any of the ensemble members is arbitrarily chosen as reference forecast for the whole verification period.
Figure \ref{fig:qav} shows the ensemble added value at specific probability levels. 
The probability levels investigated here are the ones defined by the ensemble size
($\tau=2.5\%,7.5\%,...,92.5\%,97.5\%$, see Equation \ref{equ:tau}).   
\\

The added value of the ensemble approach is statistically significantly positive all along the year. The potential benefit ranges from 4\% in the winter up to  8\% during the summer, showing a clear seasonal signal. A similar seasonal cycle is drawn  in Figure \ref{fig:qav} for intermediate probability levels. The seasonal signal is stronger for low probability levels and has an opposite behavior for high probability levels. The added value for the probability levels corresponding to the members of rank 18 to 20 exhibit 
a maximum reached during the winter and a minimum reached during the summer. The low predictability of the upper tail of the predictive distribution during the winter season and of the lower tail during the summer season can explained the clear advantage of using an ensemble system rather than a single forecast in such situations. So, the benefit of the ensemble approach for a user can vary by a factor 10 depending on the season and his or her probability level of interest. 
\\

\begin{figure*}
\centering
\includegraphics[width=14cm]{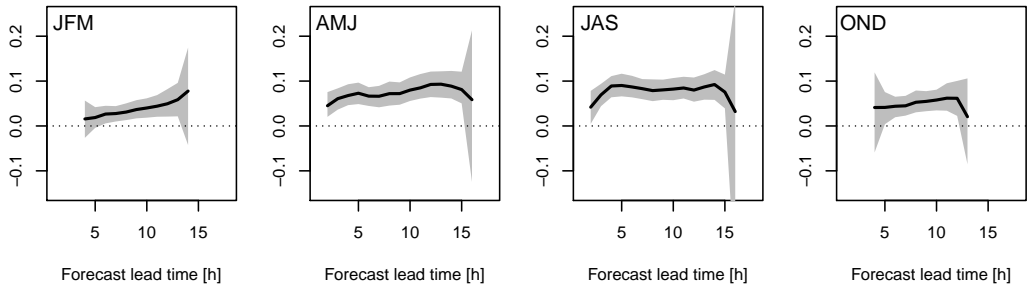}
\caption{ 
\protect{
EAV,  added value of the ensemble forecast strategy compared to a single forecast approach, as a function of the forecast lead time. From left to right, the verification period is  January - February - March (JFM), April - May - June (AMJ),  July - August - September (JAS) and October - November - December (OND) of the year 2013. The grey areas represent the 5\%-95\% confidence intervals estimated with bootstrapping. 
}
}   
\label{fig:eavt}
\end{figure*}

Finally, the analysis of the EAV was extended to all meaningful lead times (excluding the 'night' hours, see Section \ref{sec:data}). Figure \ref{fig:eavt} shows EAV as a function of the forecast horizon for four different verification periods of three months: January - February - March (JFM), April - May - June (AMJ), July - August - September (JAS) and October - November - December (OND) of the year 2013. The ensemble added value is statistically significantly greater than zero for all lead times except for 'edge' hours.
The first and last verification hours of each season, near sunset and sunrise hours, are affected by the scarcity of data that translates into large confidence intervals in the results. Otherwise, EAV is higher during the Spring/Summer than in the Autumn/Winter for all forecast horizons.  The potential benefit of using the ensemble shows also a tendency to increase as a function of the lead time for all seasons.

\section{Conclusion}
\label{sec:conc}
%the main weather parameter affecting solar production
In this study, global radiation forecasts from the high-resolution ensemble system \linebreak COSMO-DE-EPS are assessed against hourly measurements from 32 pyranometer stations. The analysis of skill scores and potential skill scores with climatology as a reference gives an estimation of the system performance over the year 2013. The comparison  of skill scores and potential skill scores highlights the deficiencies of the ensembleforecasts and derived quantile products in terms of reliability. The benefit one can expect from calibration is shown to be higher during the winter and autumn seasons and when focusing on low probability levels. The potential skill scores analysis indicates that the resolution ability of the ensemble system is relatively stable along the year, performance at low and high probability levels being balanced for each season.
\\

It is also shown that the ensemble forecasting approach is expected to be beneficial to the weather forecasts users compared to a single forecast approach. The estimation of the ensemble potential benefit is measured by a new metric called ensemble added value (EAV), aiming at a fair comparison between ensemble forecasts and single deterministic forecasts. The EAV computation is based on the decomposition of the quantile score and the ensemble interpretation of the continuous ranked probability score. EAV has the form of a skill score and rewards the additional information content provided by the ensemble forecast. For a local global radiation forecast, the  benefit gained from the ensemble approach is  statistically significant for all relevant forecasting hours and all seasons. The added value of the ensemble is greatest during spring-summer periods and increases with the lead time. 

\section*{Ackowlegment} 
This work has been realized in the framework of the EWeLiNE  project (\textit{Erstellung innovativer Wetter- und Leistungsprognosemodelle f\"ur die Netzintegration wetterabh\"angiger Energietr\"ager}) funded by the German Federal Ministry for Economic Affairs and Energy.  Tobias Heppelmann, Richard J. Keane, Kristina Lundgren and two anonymous reviewers are thanked for their valuable comments on draft versions of this manuscript.

\end{document}